# Theoretical elucidation of possibility of Majorana modes in a two dimensional Dirac system[*]

**Partha Goswami**
Physics Department, D.B.College, University of Delhi, Kalkaji, New Delhi-110019, India
Email: physicsgoswami@gmail.com

**Abstract**

Here we present the theoretical clarification of possibility of eight Majorana-like modes (quasi-particles which are self-conjugate) close to the experimentally inaccessible Dirac points of a two-dimensional monolayer Dirac system. The valley-mixing and the spin-degeneracy lifting are the main requirements. These are possible by wedging in the requisite ingredients in the description, viz. the atomically sharp scatterers and the strong spin-orbit coupling (SOC). The latter can possibly be achieved in graphene folding a sheet; the higher curvature of deformations correspond to stronger values of the coupling. In silicene, the buckled structure of the system generates a staggered sub-lattice potential between silicon atoms at A sites and B sites for an applied electric field $E_z$ perpendicular to its plane. The stronger SOC in silicene has its origin also in the buckled structure of the system. Tuning of $E_z$, allows for rich behavior varying from a topological insulator (TI)to a normal insulator (NI) with a valley spin-polarized metal (VSPM) at a critical value in between. The VSPM stage is characterized by the valley-spin locking, i.e. the opposite spin polarization at different valleys. We shall see that in this phase, if the inter-valley scattering process and the real spin-flip process in moderation are allowed to take place, we have the right condition for capturing Majoranas in the proximity of a s-wave superconductor.

## 1.Introduction

The theoretical elucidation of the possibility of Majoranas[1,2,3,4] in a novel two dimensional monolayer silicene/grapheme-like Dirac system is the goal of this communication. To explain in brief what are Majoranas, we consider the Dirac equation $i(\gamma^\mu \partial_\mu)\psi = 0$ in (1+2) dimensions for mass-less real fermions in covariant form where $\gamma^0 \equiv -i\sigma^2$, $\gamma^1 \equiv \sigma^1$, $\gamma^2 \equiv \sigma^3$, and $\sigma^j$ are Pauli matrices. In this three-component description, since the real matrices { $\gamma^0$, $\gamma^1$, $\gamma^2$ } render the Hamiltonian real, the full, space-time dependent field $\psi$ is complex. The field $\psi$ here corresponds to a mass-less Dirac fermion (or a charge-non-self-conjugate fermion). We now consider a scenario where the field $\psi$ is real and corresponds to a charge-self-conjugate Dirac fermion satisfying a Dirac equation $i(\gamma^\mu \partial_\mu)\psi = 0$ involving imaginary $\gamma$-matrices. Accordingly, one chooses a different representation for such matrices, viz. $\gamma^0 = \sigma^2$, $\gamma^1 = i\sigma^3$, $\gamma^2 = i\sigma^1$ where all the matrices are imaginary. The field $\psi$ here corresponds to a mass-less Majorana fermion. Having clarified this we note that graphene has weak intrinsic spin-orbit interaction(15- 30 meV) (as the carbon nuclei is light) and weak hyperfine coupling as carbon materials consist predominantly of the nuclear spin free $^{12}C$ isotope. Though this makes graphene a potentially a spin conductor with long spin coherence times [5,6] and opens a negligible energy gap at the Dirac points, it does not match with our goal. However, since silicon is heavier than carbon, the spin-orbit coupling in silicene is naturally much larger than in graphene. The unit cell of silicene contains two atoms which gives rise to two different sub-lattices A and B as in graphene. The honeycomb lattice of the former system, however, is distorted due to a large ionic radius of a silicon atom and forms a buckled structure pointing out-of-plane. Furthermore, the stronger SOC in silicene has its origin also in the buckled structure of the former. As we shall show in section 2 that the spin-degeneracy lifting is an important requirement for the possibility of Majoranas in these systems. This requirement poses no problem in silicene. The strong Rashba spin-orbit (RSOC), tunable by gate voltage, leads to the reduction of the gap and the spin-degeneracy lifting in graphene. It may be mentioned that one may also enlarge RSOC in graphene via doping low-concentration 3d or 5d transition metal atoms on the hollow adsorption sites[7]. Indeed, a large Rashba spin-orbit interaction had been reported by Varykhalov et al. not long ago[8]. It is also well-known[9,10] that folding of a graphene sheet gives rise to the spin-orbit coupling enhancement surrounding (non-planar) deformations.

In order to explain yet another requirement we consider the kinetic term in the single particle Hamiltonian ($H_0$) in real space of the Dirac systems, which may be represented in a compact form by $H_0 = -i\hbar\, v_F\, (\tau_3 \otimes \sigma_1 \partial_x + \tau_3 \otimes \sigma_2 \partial_y)$, where $\tau_{0,1,2,3}$ ($\tau_0 = I_{2X2}$(identity matrix), $\tau_1 = \tau_x$, $\tau_2 = \tau_y$, $\tau_3 = \tau_z$) and $\sigma_{0,1,2,3}$($\sigma_0 = I_{2X2}$, $\sigma_1 = \sigma_x$, $\sigma_2 = \sigma_y$, $\sigma_3 = \sigma_z$) are two independent sets of $2 \times 2$ Pauli matrices. In graphene, the Pauli matrices $\tau_{0,1,2,3}$ correspond to the **K** and **K′** valley (iso-spin) index whereas the Pauli matrices $\sigma_{0,1,2,3}$ correspond to the A and B sub-lattice (pseudo-spin) index. The Dirac equation with this Hamiltonian is given by $H_0 \Psi = \varepsilon \Psi$ where $\Psi = (\psi^B_K \quad \psi^A_K$ $\psi^B_{K'}$ $\psi^A_{K'}$ $)^T$ is a four component spinor. To introduce the real spin, of course, yet another grading (represented by the Pauli matrices $s_{0,1,2,3}$) must be inserted which we shall discuss in section 3. An additional mass($m(\mathbf{r})$)term in the Dirac equation above can be viewed as a bosonic field (an order parameter) generated due to the spontaneous breaking of a symmetry, such as the chiral symmetry. This may be dubbed as the Higgs mechanism in the Dirac systems in 1+2 -space-time dimensions. There could be topological defects (TD) in the order parameter as well, such as vortices [11] in a type–II superconductor. In graphene-like system, the mass order parameter could be induced, for example, by placing the system on a certain substrate where there is a difference in the potential [12], seen by the two atoms in the unit cell of graphene, which creates a charge-density wave (CDW) gap with broken chiral symmetry. The single-particle excitation spectrum, however, is particle-hole symmetric and, therefore, "hiding" the charge difference one may construct the Majorana modes out of electron and hole excitations provided one has access to a situation characterized by broken iso-spin symmetry and, of course, spin non-degeneracy[7]. We find them as the necessary conditions required for the construction of Majorana modes in our graphene-like system. The iso-spin symmetry breaking is possible if the CDW gap generating potential corresponds to atomically sharp scatterers. To explain, we quote here that Suzuura et al. [13,14,15] in a different context have suggested several years ago that, when the inter-valley scattering rate is higher than the de-coherence rate, the inter-valley particle–particle correlation function(PPCF) is enhanced leading to a conventional weak localization(WL). These authors have reported that, even in the absence of spin-orbit coupling, from the possible weak anti-localization(WAL−positive magneto-resistance beyond a critical magnetic field $B_i$) a WL(negative magneto-resistance for all possible magnetic field strength) may be obtained by a strong inter-valley scattering from the atomically sharp scatterers, while the crossovers from the latter to the former are obtained by reducing the disorder strength down to the ballistic limit [15]. In addition, it was shown that the trigonal warping inclusion in the monolayer graphene Hamiltonian [15,16] suppresses the intra-valley PPCF and, therefore, WAL as well in the case when electrons do not change their valley state; the inter-valley PPCF is not affected by trigonal warping in the case of weak inter-valley scattering due to the time-reversal invariance of the system. In view of these published results we visualize (do not visualize) a major role of the inter-valley scattering(the trigonal warping) in the search Majoranas in graphene-like system.

Alternative to the complex scenario portrayed in the paragraph above is that the substrate could be a superconductor leading to a particle-hole symmetric excitation spectrum. Moving on with this choice, we recall that the states in a Dirac system at different energy (that is states from the valence band and from the conduction band), in general, arise from the different valleys $\mathbf{K}$ and $\mathbf{K}'$. Once again, one is interested here in zero-energy ($\varepsilon = 0$) mid-gap, real solutions (Majorana-like quasi-particles) which should be localized/ quasi-localized in space. We note that in order to localize such states one needs to have TDs, such as quantum vortices [11], which can trap the "so-called" mid-gap zero modes(the topological protection of these modes is guaranteed by the Atiyah-Singer index theorem[17,18]). To pave the ground to include the superconducting order parameters (together with TDs as vortices) in the Dirac equation mentioned above, one must introduce one more grading relating to particle-hole (represented by the Pauli matrices $\mu_{0,1,2,3}(\mu_0 = I_{2X2}, \mu_1 = \mu_x, \mu_2 = \mu_y, \mu_3= \mu_z)$) in the equation. Thus, including a vector potential $\mathbf{A}= (A_x, A_y)$( equivalently, $\mathbf{A}= -\mathbf{e_\theta}A(r)$, where $\mathbf{e_\theta} =(-\sin(\theta),\cos(\theta))$, in the plane polar coordinate system), one may now write the full Hamiltonian in compact form as $H = [v_F (\tau_3 \otimes\mu_0\otimes\sigma_1$ Л$_x + \tau_3\otimes\mu_3\otimes\sigma_2$ Л$_y) + m(\mathbf{r})\otimes\mu_3 \otimes\sigma_0]$ where the operators Л$_x \equiv - i\hbar\partial_x$ –q $A_x$, Л$_y \equiv -i\hbar\partial_y$–q$A_y$, and the matrix $m(\mathbf{r}) = \begin{pmatrix} 0 & \Delta(r) \\ \Delta^*(r) & 0 \end{pmatrix}$. To introduce the real spin, of course, yet another grading (represented by the Pauli matrices $s_{0,1,2,3}$) must be inserted. The Hamiltonian H is the low-energy Ghaemi-Wilczek [19] version of the Dirac Hamiltonian with the usual Peierls substitution including the superconducting order parameter $\Delta(\mathbf{r})$. The pair potential corresponds to opposite 2-momentum spin-singlet states requiring inter-valley mixing[19]. With this type of pairing, the authors[19] could show the existence of quasi-localized near zero modes in graphene.

The paper is organized as follows: In section 1 we focus on the low-energy spectrum of a Dirac system, viz. silicene, rather than graphene. Our preference for the silicene hinges on the facts that the silicene is a better candidate due to its stronger SOC to take our discussion to the next level, and one of the most notable limitations of the graphene is its incompatibility with the existing silicon-based electronics. This is likely to be taken care of in silicene − a rival of graphene, as the former is a material relatively easy to incorporate within the existing electronics. With this we show that the broken iso-spin symmetry and the moderate presence of the magnetic impurities will gives rise to appreciably high probability of capturing Majorana modes for the system under consideration in the proximity with a conventional s-wave superconductor. In section 3 we construct Majorana operators in terms of Dirac creation and annihilation oper-ators of second quantization. The paper ends with some concluding remarks in section 4.

## 2. Requirements of hosting Majorana in silicene

The discussion in the previous section clearly indicates that the Majoranas and the Dirac fermions in a 1+2 space-time system are intimately connected. The main issue from a theoretical perspective is to find the condition(s) under which the former could be realized in such systems. To look for the condition(s) in a Dirac system, we shall express mathematically the Dirac and Majorana fermions in terms of the creation and annihilation operators of second quantization in this section. Our expressions will encode the electron's and hole's characteristic fermion statistics, the particle–antiparticle correspondence, and the unusual anti-commutation relations obeyed by the Majorana operators in a transparent manner. Before taking up this task we note that our candidate [18] is an operator $\hat{C}_j = d_j^\dagger + d_j$ which corresponds to a particle and hole mixture creation in a state j. Here $d_j^\dagger$ can create a particle, or destroy a hole in a state j, whereas $d_j$ can create a hole, or destroy a particle, in a state j. For two distinct (orthogonal) states j and k, the anti-symmetry of Fermi-Dirac statistics implies that $\{d_j^\dagger, d_k^\dagger\} = \{d_j, d_k^\dagger\} = \{d_j, d_k\} = 0$. The completeness relation, on the other hand, implies that $\{d_j, d_j^\dagger\} = 1$. We notice that the particle–hole interchange (charge conjugation) is implementable by $d_j \leftrightarrow d_j^\dagger$. Thus, for a Dirac fermion the operators $d_j^\dagger$ and $d_j$ are distinct, while for a Majorana fermion they are identical. One example, which complies apparently with the implementation, is an exciton (bound states of electron and hole). In the language of second quantization, this will correspond to $\hat{A}_{exciton} \equiv (d_j d_k^\dagger + d_j^\dagger d_k)$. Obviously enough, under charge conjugation, the exciton 'creation' operator $\hat{A}_{exciton}$ goes over to itself, and the concomitant excitations are their own antiparticles. But excitons are always bosons, with integer spin, and thus could not be Majoranas. Our candidate, therefore, is an operator $\hat{C}_j = d_j^\dagger + d_j$ which corresponds to a particle and hole mixture creation in a state j. It will hide the 'charge' completely without tinkering with the spin.

We now consider the monolayer silicene as a better candidate due to its stronger SOC to take our discussion to the next level. The Kane-Mele Hamiltonian [20] of the system, including both intrinsic and Rashba spin-orbit coupling, can be written in the following form. The dimensionless Hamitonian matrix around Dirac point $K_\xi$ (the iso-spin index $\xi = \pm 1$) in the basis ($a_{\mathbf{k}\uparrow}$, $b_{\mathbf{k}\uparrow}$, $a_{\mathbf{k}\downarrow}$, $b_{\mathbf{k}\downarrow}$) in momentum space is h($\delta\mathbf{k}$)=[$\xi a\, \delta k_x (\gamma^5\gamma^0\gamma^x)$ + a $\delta k_y (\gamma^5\gamma^0\gamma^y)$ ] + $\xi[t'_{soc}(\gamma^5\gamma^z\gamma^0\gamma^5) + \Delta_z(\gamma^5\gamma^z\gamma^0)$ +$ai$ $t'_{Rashba}(\gamma^z\delta k_x + \gamma^5\gamma^z\delta k_y)$]–$M\gamma^5$. Here 4X4 matrices($\gamma$) are in chiral basis. The first term is the kinetic energy. In a tight-binding approximation, the central term ($-\Delta_z \sum_{i,\sigma} \mu_i\ c^\dagger_{i\sigma} c_{i\sigma}$) with **ℓ = 0.23Å**, $\Delta_z = \ell E'_z$ ($E_z$ is the electric field) is the staggered sub-lattice potential term where $\mu_i = \pm 1$ for the A(B) site. These terms break the sub-lattice symmetry of the silicene's honey-comb structure and generate a gap. The exchange field **M** may arise due to coupling to a ferromagnet (FM) such as depositing Fe atoms to the silicene surface or depositing silicene to an FM insulating substrate. The terms $t'_{soc}(\gamma^5\gamma^z\gamma^0\gamma^5)$ and $ai$ $t'_{Rashba}(\gamma^z\delta k_x + \gamma^5\gamma^z\delta k_y)$ correspond to spin-orbit coupling. In the Kane-Mele framework, the corresponding terms are ($H_{SOC} + H_{RSOC}$): The term $H_{SOC}$ is the matrix $\begin{pmatrix} \xi\Delta\sigma_z & 0 \\ 0 & -\xi\Delta\sigma_z \end{pmatrix}$ and $H_{RSOC}$ is $\begin{pmatrix} 0 & \lambda(\xi\sigma_y + i\sigma_x) \\ \lambda(\xi\sigma_y - i\sigma_x) & 0 \end{pmatrix}$. The $\Delta$ and $\lambda$, respectively, are the parameters specifying the intrinsic and Rashba spin-orbit couplings. As already mentioned above, for graphene, it is possible[9,10] that folding of a sheet gives rise to the spin-orbit coupling enhancement surrounding (non-planar) deformations. Following refs. [21, 22], one knows that $\lambda$ is directly proportional to the curvature of the deformations, which means that higher curvatures correspond to stronger values for $\lambda$. The eigenvalues($\varepsilon$) of the Kane-Mele matrix are given by $\varepsilon = \pm\sqrt{[(\hbar v_F |\mathbf{k}|)^2 + \Delta^2 + 2\lambda^2 \pm 2\sqrt{\{\lambda^4 + (\hbar v_F |\mathbf{k}|)^2 \times (\Delta^2 + \lambda^2)\}}]}$.

In the absence of the exchange field and intrinsic Rashba originating from the buckled honey-comb structure, we find the following bands from h($\delta\mathbf{k}$): $\mathcal{E}(\delta k) = \pm[(a|\delta\mathbf{k}|)^2 +\{\xi\, s_z\, \Delta_{soc} + \Delta_z\}^2]^{1/2}$ where $\Delta_{soc} = t'_{so}$, $s_z = \pm 1$ for $\{\uparrow, \downarrow\}$. The effective staggered sub-lattice potential V= $\{\xi\, s_z\, \Delta_{soc} + \Delta_z\}$. The time reversal symmetry (TRS) requires $\mathcal{E}(\xi, s_z\, \delta\mathbf{k}) = \mathcal{E}(\xi, s_z\, -\delta\mathbf{k})$. The low-energy spectrum given above also comes up from the matrix

$$\hbar(\delta\mathbf{k},) = [\xi a\tau^0 \otimes \sigma^x\ \delta k_x + a\tau^0 \otimes \sigma^y\ \delta k_y + \xi\, \Delta_{\mathbf{soc}}\tau^z \otimes \sigma^z$$
$$+ \Delta_{\mathbf{z}}\,\tau^0 \otimes \sigma^z - M\,\sigma^0 \otimes \tau^z - (\mu)\,\tau^0 \otimes \sigma^0] \quad (1)$$

where $\tau^i$ and $\sigma^i$, respectively, denote the Pauli matrices associated with the real spin and pseudo-spin of the Dirac electronic states. The basis chosen is ($a_{\delta\mathbf{k}\uparrow}$, $b_{\delta\mathbf{k}\uparrow}$, $a_{\delta\mathbf{k}\downarrow}$, $b_{\delta\mathbf{k}\downarrow}$). In writing this Hamiltonian we have ignored the intrinsic Rashba terms $\sum_{\delta\mathbf{k}} \{\xi a(\delta k_y + i\delta k_x)\, a^\dagger_{\delta\mathbf{k},\uparrow} a_{\delta\mathbf{k},\downarrow} + (\delta k_y - i\delta k_x)\, b^\dagger_{\delta\mathbf{k},\uparrow} b_{\delta\mathbf{k},\downarrow}$ + h.c.$\}$ as $t_2 \ll \Delta_{SOC}$. No binding of transition metal molecules to the surface have been assumed. The TM-induced stronger extrinsic Rashba SOC leads to spin-splitting[23].

For comparison, at the Γ point of the surface state Brillouin zone of a topological insulator (TI) we write the surface state Hamiltonian $\mathcal{K}(\delta k) = \varepsilon(\delta k) + \hbar v(\delta k)(\delta k_x \sigma^y - \delta k_y \sigma^x) + (\lambda/2)(\delta k_+^3 + \delta k_-^3)\sigma^z$ where $\sigma^i$ denote the Pauli matrices associated with the spin of the Dirac electronic states. There is one important difference between these two cases. For silicene, the two components of the Dirac Hamiltonian describe the two sub-lattices or pseudo-spin degrees of freedom, while in the case of a TI, the two components describe the real electron spin, and are related to each other by time reversal.

In view of (1), it is possible to write a Hamiltonian where only the pseudo-spin is in the foreground; the iso-spin(described by the index $\xi = \pm 1$) and the real spin (described by an index $s_z = \pm 1$) are in the background. Such a phenomenological, minimal Hamiltonian matrix for the ferromagnetic silicene could be written as
.

$\hbar_{reduced}(\xi, s_z, \delta \mathbf{k})/\left(\frac{\hbar v_F}{a}\right) \approx \sum_{\delta \mathbf{k}, s_z}[(\xi\, s_z \Delta_{soc} + \Delta_z - s_z M)\, a^\dagger_{\delta \mathbf{k}, s_z} a_{\delta \mathbf{k}, s_z} + \{\xi\, a\, \delta k_x - ai\, \delta k_y\}\, a^\dagger_{\delta \mathbf{k}, s_z} b_{\delta \mathbf{k}, s_z} + (-\xi\, s_z \Delta_{soc} - \Delta_z - s_z M)\, b^\dagger_{\delta \mathbf{k}, s_z} b_{\delta \mathbf{k}, s_z} + \{\xi\, a\, \delta k_x + ai\, \delta k_y\}\, b^\dagger_{\delta \mathbf{k}, s_z} a_{\delta \mathbf{k}, s_z} - (\mu/\left(\frac{\hbar v_F}{a}\right))(a^\dagger_{\delta \mathbf{k}, s_z} a_{\delta \mathbf{k}, s_z} + b^\dagger_{\delta \mathbf{k}, s_z} b_{\delta \mathbf{k}, s_z})]$. (2)

In the absence of the non-magnetic impurities, the bands are given by $\mathcal{E}(\delta k) = -s_z M \pm [(a|\delta \mathbf{k}|)^2 + \{\Delta_{soc} + \xi\, s_z \Delta_z\}^2]^{1/2} - \mu'$. Here $\mu' = (\mu/\left(\frac{\hbar v_F}{a}\right))$ is the dimensionless chemical potential of the fermion number. The corresponding eigenvectors are now two-component Dirac spinor:

$|\gamma\rangle = \begin{pmatrix} \left(\frac{\hbar v_F}{a}\right) ak'_+ \\ E_M \end{pmatrix}$, $|\delta\rangle = \begin{pmatrix} -\left(\frac{\hbar v_F}{a}\right) ak'_- \\ E_M \end{pmatrix}$, where

$E_M(\xi, s_z, a|\mathbf{k}'|) = E(a|\mathbf{k}'|) + \Delta_{soc}^{(M)}(\xi, s_z, a|\mathbf{k}'|) + s_z M$,

$E = \pm[\{(a|\mathbf{k}'|)^2 + \Delta_{soc}^{(M)}(\xi, s_z, a|\mathbf{k}'|)^2\}]^{1/2} - s_z M$,

$\Delta_{soc}^{(M)}(\xi, s_z, a|\mathbf{k}'|) \equiv (\xi s_z \Delta_{soc}(a|\mathbf{k}'|) + \Delta_z)$,

$(ak'_x) = [(E(a|\mathbf{k}'|) + s_z M)^2 - \Delta_{soc}^{(M)}(\xi, s_z, a|\mathbf{k}'|)^2 - (ak_y)^2]^{1/2}$, $k'_\pm = \xi\, k'_x \pm i\, k_y$. $\Delta_{soc}(a|\mathbf{k}'|) = (t'^2_{so} + (at'_2|\delta \mathbf{k}|)^2)^{1/2}$. (3)

It is, thus, possible to describe electrons by the effective two-component wave function. It is also possible to calculate almost all the properties of silicene with this description.

We now consider the normal silicene and continue with this system only. Here from (2), for example, for $s_z = -1$ and $\xi = +1$, one has gap closing, i.e.

$E(\mathbf{k}, s_z = -1, \xi = +1)$

$\approx \pm[\{(a|\mathbf{k}|)^2 + (\,(t'^2_{so} + (at'_2|\delta \mathbf{k}|)^2)^{1/2} - \Delta_z)^2\}]^{1/2} \quad -\mu'$

$\approx \pm(a|\mathbf{k}|)$

for $\mu' = 0$ due to $\Delta_{zc} = \Delta_z \approx (t'^2_{so} + (at'_2|\delta \mathbf{k}|)^2)^{1/2}$. However, in this case at the other $\mathbf{K}$ point $E(\mathbf{k}, s_z = -1, \xi = -1) \approx \pm[\{(a|\mathbf{k}|)^2 + 4\Delta_z^2\}]^{1/2}$ is gapped. One also finds that the other spin band is not gapped($E_{renorm}(\mathbf{k}, s_z = +1, \xi = -1) \approx \pm(a|\mathbf{k}|)$ for $\mu' = 0$) for $\mathbf{K}'$ and gapped($E_{renorm}(\mathbf{k}, s_z = +1, \xi = +1) \approx \pm[\{(a|\mathbf{k}|)^2 + 4\Delta_z^2\}]^{1/2}$) for $\mathbf{K}$. The preliminary elucidation of the requirements for hosting Majoranas in our system is the task at hand now. This necessitates us to go back to the valley-spin-polarized metal(VSPM)case [29,30, 31,32,33].

Ever since Fu and Kane have predicted [25] a one-dimensional mode of Majorana fermions (half-integer-spin (relativistic) particles which are their own anti-particles) at the interface between a conventional super-conductor and a superconducting topological insulator surface state, there has been persistent effort[26,27,28] to obtain signature of this elusive mode within such systems. In the backdrop of the excitement generated due to this finding, perhaps it is useful to discuss at the very preliminary level how such modes could be hosted/ captured in the present 2D Dirac system in the proximity of a conventional s-wave superconductor. It may be noted that the Majorana modes were originally predicted by E. Majorana[4] nearly seventy years ago in 1+3 space-time dimensions.

Coming back to the system at hand, we note that the iso-spin non-conserving processes are must for hosting Majoranas as momentum states in conduction and valence bands can be associated with the both the valleys $\mathbf{K}$ and $\mathbf{K}'$ and the band operators are needed for writing down the Majorana operators in the second quantized notations. Now if there are mass-less Dirac and Majorana fermions of certain spin variety, say spin-up state, residing at $\mathbf{K}$, after inter-valley scattering these particles transform into the massive Dirac particles of the same spin variety at $\mathbf{K}'$ rendering them elusive. The Dirac and Majorana fermions of opposite spin variety at $\mathbf{K}'$ get transformed in a similar manner at $\mathbf{K}$ upon undergoing the scattering. This scenario is evidently non-conducive for hosting Majoranas. We are, however, also able to see from the expressions of $E(\mathbf{k}, s_z = -1, \xi = +1)$, $E(\mathbf{k}, s_z = +1, \xi = +1)$ etc., that the real spin-flip process keeps the pseudo-spin unchanged when the inter-valley scattering takes place. The probability of such a spontaneous spin flip is, admittedly, low in the absence of magnetic impurities or magnetic fields. Therefore, a moderate dose

## 3. Construction of Majorana operators in terms of Dirac creation and annihilation operators of second quantization

We now turn our attention to a basic Dirac Hamiltonian $\hbar_0(\delta \mathbf{k}) = [\xi a\tau^0 \otimes \sigma^x \delta k_x + a\tau^0 \otimes \sigma^y \delta k_y]$. This Hamiltonian describes the $(\mathbf{k}, s_z = -1, \xi = +1)$ and $(\mathbf{k}, s_z = +1, \xi = -1)$ states of the system in the VSPM phase. The remaining states correspond to the massive Dirac fermion in the VSPM stage. The matrices $\hbar_{0\mathbf{K}}(\delta \mathbf{k})$ and $\hbar_{0\mathbf{K}'}(\delta \mathbf{k})$ have two eigenvalues $\pm \hbar v_F |\delta \mathbf{k}|$. The normalized eigenfunction may be written as

$$\psi_{\pm, \mathbf{K}}(\delta \mathbf{k}, s_z = -1) = (1/\sqrt{2}) \begin{pmatrix} \exp(-\left(\frac{1}{2}\right) i\theta_k) \\ \pm \exp(\left(\frac{1}{2}\right) i\theta_k) \end{pmatrix},$$

$$\psi_{\pm, \mathbf{K}'}(\delta \mathbf{k}, s_z = -1) = (1/\sqrt{2}) \begin{pmatrix} \exp(\left(\frac{1}{2}\right) i\theta_k) \\ \mp \exp(-\left(\frac{1}{2}\right) i\theta_k) \end{pmatrix}, \quad (4)$$

where $\cos(\theta_k) = \delta k_x / |\delta \mathbf{k}|$, $\sin(\theta_k) = \delta k_y / |\delta \mathbf{k}|$, and $\theta_k = \arctan(\delta k_y / \delta k_x)$. In what follows we present a general discussion without the explicit book-keeping of the spin state $s_z = \pm 1$. We have $+\hbar v_F |\delta \mathbf{k}| \rightarrow \psi_{+, \mathbf{K}(\mathbf{K}')}(\delta \mathbf{k})$ (conduction band(electron state)) and $-\hbar v_F |\delta \mathbf{k}| \rightarrow \psi_{-, \mathbf{K}(\mathbf{K}')}(\delta \mathbf{k})$ (valence band(hole state)). One notices that $\psi_{+,\mathbf{K}}$ is complex conjugate of $\psi_{-,\mathbf{K}'}$ and $\psi_{-,\mathbf{K}}$ is complex conjugate of $\psi_{+,\mathbf{K}'}$. We make the identification that $\psi_{c, \mathbf{K}(\mathbf{K}')}(\delta \mathbf{k}) = \psi_{+, \mathbf{K}(\mathbf{K}')}(\delta \mathbf{k})$ and $\psi_{v, \mathbf{K}(\mathbf{K}')}(\delta \mathbf{k}) = \psi_{-, \mathbf{K}(\mathbf{K}')}(\delta \mathbf{k})$ where the subscript c(v) refers to the conduction(valence) band. We find that $\psi_{c(v), \mathbf{K}}(\delta \mathbf{k}) = (1/\sqrt{2})(|1\rangle_\mathbf{K} \pm |0\rangle_\mathbf{K})$, and $\psi_{c(v), \mathbf{K}'}(\delta \mathbf{k}) = (1/\sqrt{2})(|1\rangle_{\mathbf{K}'} \mp |0\rangle_{\mathbf{K}'})$, where the upper(lower) sign corresponds to the subscript c(v). We have $|1\rangle_{\mathbf{K}(\mathbf{K}')} = \exp(\mp i\theta_k/2) |\uparrow\rangle$ and $|0\rangle_{\mathbf{K}(\mathbf{K}')} = \exp(\pm i\theta_k/2) |\downarrow\rangle$; in the states $|1\rangle_{\mathbf{K}(\mathbf{K}')}$ and $|0\rangle_{\mathbf{K}(\mathbf{K}')}$ the upper(lower) sign corresponds to the subscript $\mathbf{K}(\mathbf{K}')$. Here $|\uparrow\rangle = \begin{bmatrix} 1 \\ 0 \end{bmatrix}$ and $|\downarrow\rangle = \begin{bmatrix} 0 \\ 1 \end{bmatrix}$, respectively, are the 'so-called' up and down states arising out of the choice of the sub-lattice basis (A, B). We, therefore, notice that the existence of two independent sub-lattices A and B(corresponding to the two atoms per unit cell) leads to the existence of novelty in graphene/silicene dynamics where the two linear branches of energy dispersion (intersecting at Dirac points) become independent of each other, indicating the existence of a pseudo-spin quantum number analogous to electron spin (but completely independent of real spin). In other words, the existence of the pseudo-spin quantum number is a natural byproduct of the basic lattice structure of graphene/silicene comprising two independent sub-lattices. The eigenstates above in the vicinity of the $\mathbf{K}$ and $\mathbf{K}'$ points necessitate the introduction of the notion of iso-spin, once again reminiscent of the states of the spin-1/2 operator.

To clarify the notion of the iso-spin, we note that $\psi_{\pm, \mathbf{K}}(\delta \mathbf{k})$ and $\psi_{\pm, \mathbf{K}'}(\delta \mathbf{k})$ are linked by a symmetry property provided we establish the correspondence between the states around the valleys $\mathbf{K}$ and $\mathbf{K}'$ with real single spin-1/2 operator. For this we draw the analogy with a single real spin-1/2 operator $\mathbf{S}$ represented in terms of Pauli matrices $\sigma^i$: $S^i = (1/2)\hbar \sigma^i$. The eigenvalues of $\sigma^i$ are $\pm 1$ and the corresponding eigenstates of $\sigma^z$, say, are $|\uparrow\rangle = \begin{pmatrix} 1 \\ 0 \end{pmatrix}$, and $|\downarrow\rangle = \begin{pmatrix} 0 \\ 1 \end{pmatrix}$. The operator $S^i$ in the second quantized language can be written as $\mathbf{S}^i = \sum_{\mu,\mu'} d^\dagger_\mu S^i_{\mu,\mu'} d_{\mu'}$ where $d^\dagger_\mu$ creates a particle in the state $|\mu\rangle$. This immediately gives $S^x = (1/2)(d^\dagger_\uparrow d_\downarrow + d^\dagger_\downarrow d_\uparrow)$, $S^y = (1/2i)(d^\dagger_\uparrow d_\downarrow - d^\dagger_\downarrow d_\uparrow)$, and $S^z = (1/2)(d^\dagger_\uparrow d_\uparrow - d^\dagger_\downarrow d_\downarrow)$. The spin-reversal operators are $S^+ = d^\dagger_\uparrow d_\downarrow$ and $S^- = d^\dagger_\downarrow d_\uparrow$. The anti-unitary time reversal operator for real spins is defined as $\hat{A} = \Theta \kappa$ where $\Theta = \exp(i\pi S^y/\hbar)$, and $\kappa$ is the complex conjugation operator. The operator $\Theta$(an orthogonal matrix) is given by $\begin{pmatrix} 0 & 1 \\ -1 & 0 \end{pmatrix}$. One may then write $\hat{A} \psi_\uparrow = \psi^*_\downarrow$ and $\hat{A} \psi_\downarrow = -\psi^*_\uparrow$. Having accomplished this exercise, we notice that analogously $\hat{A} \psi_{+, \mathbf{K}}(\delta \mathbf{k}) = \psi_{-, \mathbf{K}}(\delta \mathbf{k}) = \psi^*_{+, \mathbf{K}'}$ and $\hat{A} \psi_{-, \mathbf{K}}(\delta \mathbf{k}) = -\psi_{+, \mathbf{K}}(\delta \mathbf{k}) = -\psi^*_{-, \mathbf{K}'}$. In fact, we also notice from above that $\hat{A} \Theta^{-1} = I$, $\hat{A} H_\mathbf{K} = -H_\mathbf{K} \Theta$ and $\hat{A} H_{\mathbf{K}'} = -H_{\mathbf{K}'} \Theta$ which yield $\Theta \kappa H_\mathbf{K} \Theta^{-1} = -H_\mathbf{K} = H^*_{\mathbf{K}'}$, or, $\Theta H^*_\mathbf{K} \Theta^{-1} = H^*_{\mathbf{K}'}$. The difference, however, is that whereas $\psi_\uparrow$ and $\psi_\downarrow$ are spinors $\begin{bmatrix} 1 \\ 0 \end{bmatrix}$ and $\begin{bmatrix} 0 \\ 1 \end{bmatrix}$ with real elements, $\psi_{+, \mathbf{K}}(\delta \mathbf{k})$, $\psi_{+, \mathbf{K}'}$, etc. are Dirac spinors with complex elements. None-the-less, upon assuming that the valley states somehow correspond to real spins we find that states around $\mathbf{K}$ and $\mathbf{K}'$ are linked by a symmetry akin to the time-reversal symmetry of real spins. In silicene/graphene, electronic density is usually shared between A and B sub-lattices, so that an iso-spin indexed wave function is a linear combination of 'up' and 'down' as shown above. We see that not only do the electrons possess the iso-spin degree of freedom, but they are chiral, meaning the orientation of the pseudo-spin $\boldsymbol{\sigma}$ is related to the direction of the electronic momentum $\mathbf{p}$. We introduce the chirality (helicity) for the system to characterize the eigenfunctions through the projection of the momentum operator along the direction of the operator $\boldsymbol{\sigma} = (\sigma_x,$ $\sigma_y$) (or $\boldsymbol{\sigma}^* = (-\sigma_x, \sigma_y)$). The chirality operator is defined as $\hat{C} = (1/2)\,\boldsymbol{\sigma} \cdot \frac{\delta \mathbf{k}}{|\delta \mathbf{k}|}$ for momentum around $\mathbf{K}$ and $\hat{C}^* = (1/2)\,\boldsymbol{\sigma}^* \cdot \frac{\delta \mathbf{k}}{|\delta \mathbf{k}|}$ for momentum around $\mathbf{K}'$.

The stage is now set to introduce the Majorana-like operators in the second quantization language for graphene. In a bid to achieve this, we first recall that the operators $a^\dagger_{\delta\mathbf{k},\sigma}$ and $b^\dagger_{\delta\mathbf{k},\sigma}$ with momentum $\delta\mathbf{k}$ and spin $\sigma$, respectively, could be used for the fermion creation operators for A and B sub-lattices. Suppose the creation operators for c(v) be denoted by $d^\dagger_{c,\delta\mathbf{k},\sigma}$ ($d^\dagger_{v,\delta\mathbf{k},\sigma}$). In view of above, around $\mathbf{K}$ we may define

$$d^\dagger_{c(v),\delta\mathbf{k},\sigma}(\mathbf{K}) = (1/\sqrt{2})\,a^\dagger_{\delta\mathbf{k},\sigma}(\mathbf{K})\exp(-i\theta_k/2)$$
$$\pm (1/\sqrt{2})\,b^\dagger_{\delta\mathbf{k},\sigma}(\mathbf{K})\exp(+i\theta_k/2).$$

This leads to

$a^\dagger_{\delta\mathbf{k},\sigma}(\mathbf{K}) = (1/\sqrt{2})\exp(i\theta_k/2)(d^\dagger_{c,\delta\mathbf{k},\sigma}(\mathbf{K}) + d^\dagger_{v,\delta\mathbf{k},\sigma}(\mathbf{K}))$,

and

$b^\dagger_{\delta\mathbf{k},\sigma}(\mathbf{K}) = (1/\sqrt{2})\exp(-i\theta_k/2)(d^\dagger_{c,\delta\mathbf{k},\sigma}(\mathbf{K}) - d^\dagger_{v,\delta\mathbf{k},\sigma}(\mathbf{K}))$.

Similarly, around $\mathbf{K}'$, we may define

$$d^\dagger_{c(v),\delta\mathbf{k},\sigma}(\mathbf{K}') = (i/\sqrt{2})\,a^\dagger_{\delta\mathbf{k},\sigma}(\mathbf{K}')\exp(i\theta_k/2)$$
$$\pm (-i/\sqrt{2})\,b^\dagger_{\delta\mathbf{k},\sigma}(\mathbf{K}')\exp(-i\theta_k/2)$$

which leads to

$a^\dagger_{\delta\mathbf{k},\sigma}(\mathbf{K}') = (-i/\sqrt{2})\exp(-i\theta_k/2)(d^\dagger_{c,\delta\mathbf{k},\sigma}(\mathbf{K}') + d^\dagger_{v,\delta\mathbf{k},\sigma}(\mathbf{K}'))$,

and

$b^\dagger_{\delta\mathbf{k},\sigma}(\mathbf{K}') = (i/\sqrt{2})\exp(i\theta_k/2)(d^\dagger_{c,\delta\mathbf{k},\sigma}(\mathbf{K}') - d^\dagger_{v,\delta\mathbf{k},\sigma}(\mathbf{K}'))$.

It is easy to see that the band operators around $\mathbf{K}$ and $\mathbf{K}'$ introduced above anti-commute. Also, around $\mathbf{K}$ and $\mathbf{K}'$, in terms of these band operators, the Hamiltonian $\hbar_0(\delta\mathbf{k})$ is given by $\hbar_0 = \sum_{\delta\mathbf{k}} \hbar v_F |\delta\mathbf{k}|\,(d^\dagger_{c,\delta\mathbf{k},\sigma} d_{c,\delta\mathbf{k},\sigma} - d^\dagger_{v,\delta\mathbf{k},\sigma} d_{v,\delta\mathbf{k},\sigma})$. We make the following combination of the band operators:

$$\gamma_{1,A}(\delta\mathbf{k}) = d^\dagger_{c,\delta\mathbf{k},\sigma}(\mathbf{K}) + i\,d_{v,\delta\mathbf{k},\sigma}(\mathbf{K}'),$$
$$\gamma_{2,A}(\delta\mathbf{k}) = d_{c,\delta\mathbf{k},\sigma}(\mathbf{K}') - i\,d^\dagger_{v,\delta\mathbf{k},\sigma}(\mathbf{K}),$$
$$\gamma_{1,B}(\delta\mathbf{k}) = d_{c,\delta\mathbf{k},\sigma}(\mathbf{K}) - i\,d^\dagger_{v,\delta\mathbf{k},\sigma}(\mathbf{K}'),$$
$$\gamma_{2,B}(\delta\mathbf{k}) = d^\dagger_{c,\delta\mathbf{k},\sigma}(\mathbf{K}') + i\,d_{v,\delta\mathbf{k},\sigma}(\mathbf{K}).$$

Going back to the fermion operators for A and B sub-lattices, we find that these combinations yield

$$(1/\sqrt{2})\,(\gamma_{1,A}(\delta\mathbf{k}) + i\,\gamma_{2,A}(\delta\mathbf{k}))$$
$$= (a^\dagger_{\delta\mathbf{k},\sigma}(\mathbf{K}) + a_{\delta\mathbf{k},\sigma}(\mathbf{K}'))\,\exp(-i\theta_k/2),$$
$$(1/\sqrt{2})\,(\gamma_{1,B}(\delta\mathbf{k}) + i\,\gamma_{2,B}(\delta\mathbf{k}))$$
$$= (b^\dagger_{\delta\mathbf{k},\sigma}(\mathbf{K}') + b_{\delta\mathbf{k},\sigma}(\mathbf{K}))\,\exp(-i\theta_k/2). \quad (4)$$

We have the sub-lattice specific four Dirac particle-hole creation operators $\hat{A}_{ph,\sigma} = (a^\dagger_{0,\sigma} + a_{0,-\sigma})$ and $\hat{C}_{ph,\sigma} = (b^\dagger_{0,\sigma} + b_{0,-\sigma})$, respectively, equal to $(1/\sqrt{2})(\gamma_{1,A,\sigma} + i\gamma_{2,A,\sigma})$ and $(1/\sqrt{2})(\gamma_{1,B,\sigma} + i\gamma_{2,B,\sigma})$ at the Fermi level, where $\gamma_{1,A,\sigma} = (1/\sqrt{2})\,(a^\dagger_{0,\sigma} + a_{0,\sigma} + a^\dagger_{0,-\sigma} + a_{0,-\sigma})$, $\gamma_{2,A,\sigma} = (1/i\sqrt{2})\,(a^\dagger_{0,\sigma} - a_{0,\sigma} - a^\dagger_{0,-\sigma} + a_{0,-\sigma})$, $\gamma_{1,B,\sigma} = (1/\sqrt{2})\,(b^\dagger_{0,\sigma} + b_{0,\sigma} + b^\dagger_{0,-\sigma} + b_{0,-\sigma})$, $\gamma_{2,B,\sigma} = (1/i\sqrt{2})\,(b^\dagger_{0,\sigma} - b_{0,\sigma} - b^\dagger_{0,-\sigma} + b_{0,-\sigma})$. The real and imaginary parts of the ordinary fermion operators $\hat{A}_{ph,\sigma}$ and $\hat{C}_{ph,\sigma}$ correspond to eight Majorana fermions as we have $\sigma = \uparrow,\downarrow$ (real spin 'up' and 'down') with $\gamma_{i,\alpha,\sigma} = \gamma^\dagger_{i,\alpha,\sigma}$ (self-conjugate) where $\alpha$ = A/B. The formal manipulations presented above shows that for the Majorana pairs to be realized it is necessary that, apart from the broken iso-spin symmetry, the spin-degeneracy should be lifted. This is carried out by SOC. The Dirac operators $\hat{A}_{ph,\sigma}$ and $\hat{C}_{ph,\sigma}$ obey the usual anti-commutation relations: $\{\hat{A}_{ph,\sigma}, \hat{A}^\dagger_{ph,\sigma'}\} = 2\delta_{\sigma\sigma'}$, $\{\hat{C}_{ph,\sigma}, \hat{C}^\dagger_{ph,\sigma'}\} = 2\delta_{\sigma\sigma'}$, $\{\hat{A}_{ph,\sigma}, \hat{A}_{ph,\sigma'}\} = 0$, and $\{\hat{C}_{ph,\sigma}, \hat{C}_{ph,\sigma'}\} = 0$. However, the Majorana operators obey unusual (the product $\gamma_{i,\alpha,\sigma}^2 = 1$ and does not vanish) anti-commutation relations: $\gamma_{i,\alpha,\sigma}\,\gamma_{j,\beta,\sigma'} + \gamma_{j,\beta,\sigma'}\,\gamma_{i,\alpha} = 2\,\delta_{ij}\,\delta_{\alpha\beta}\,\delta_{\sigma\sigma'}$.

Unlike real $\gamma$-matrices for Dirac fermions, in 1+2 space-time dimensions there exist two in-equivalent representations for complex $\gamma$-matrices characterizing Majoranas: $\gamma^0 = \sigma^2$, $\gamma^1 = i\sigma^3$, $\gamma^2 = i\xi\sigma^1$ (where $\xi = \pm 1$). We have mentioned this in section 1. We normally use the first of these representations [35] for the expansion around the Dirac point $\mathbf{K}$ and the second one for the point $\mathbf{K}'$. It is now easy to see that the velocity operator for Majoranas is $v_F(\gamma^2 = i\xi\sigma^1, \gamma^0 = \sigma^2)$.

## 4. Concluding Remarks

We note that our scheme to realize Majorana fermions by using spin-orbit interaction (SO) (and Zeeman magnetic field) is not entirely a novel one. In a different situation - a BCS s-wave super-fluid of ultra-cold fermionic atoms in an optical lattice with a laser-field-generated effective

SO interaction- was considered for the first time by Sato et al.[1,2].Subsequently, in all cold atom physics related investigations, SOC had been realized using atom-laser coupling [36-40].This progress offered an opportunity to realize and manipulate Majorana fermions in a highly controllable manner. Sato et al.[1,2] have derived an important condition $h > \sqrt{\mu^2 + \delta^2}$ for the Majorana fermion for the first time, where h is the Zeeman field, μ is the chemical potential, and δ is an s-wave gap function. The same scheme was subsequently considered by Sau et al.[3] in a different setting. Actually the model Hamiltonian of Sau et al.[3] is identical to that in the second work in reference [1,2]. Under the condition $h > \sqrt{\mu^2 + \delta^2}$, it was shown by Sato et al.[6] that the bulk topological number of the system considered becomes nonzero, a topologically protected Majorana edge mode appears, and a Majorana zero mode exists in a vortex core.

In section 3 we have seen that around **K** and **K′** the diagonalized Hamiltonian $H_0$, in terms of the band operators together with a staggered a potential ( which may be written in the form '$mv_F^2$')which takes on different values on the two sub-lattices, appears as

$\sum_{\delta k} \hbar v_F |\delta \mathbf{k}| (d^\dagger_{c,\delta k,\sigma} d_{c,\delta k,\sigma} - d^\dagger_{v,\delta k,\sigma} d_{v,\delta k,\sigma})$

$+ \sum_{\delta k} m v_F^2 (d^\dagger_{c,\delta k,\sigma} d_{c,\delta k,\sigma} + d^\dagger_{v,\delta k,\sigma} d_{v,\delta k,\sigma})$.

Since a momentum (δ**k**) state in 'c' and 'v' can either be associated with the valley **K** (with probability amplitudes,respectively, as $\alpha_K$ and $\beta_K$)or with the valley **K′** (with probability amplitudes,respectively, as $\alpha_{K'}$ and $\beta_{K'}$), one may write the most general momentum (δ**k**) state in 'c' as $|\Psi(\delta \mathbf{k})\rangle_c = (\alpha_K \psi_{c,K}(\delta \mathbf{k}) + \alpha_{K'} \psi_{c,K'}(\delta \mathbf{k}))$. Similarly, the most general momentum (δ**k**) state in 'v' is $|\Psi(\delta \mathbf{k})\rangle_v = (\beta_K \psi_{v,K}(\delta \mathbf{k}) + \beta_{K'} \psi_{v,K'}(\delta \mathbf{k}))$. A real-space spinorial state $\psi(\mathbf{r})$ is given by $\psi(\mathbf{r}) = (\Omega_{BZ})^{-1} \int \delta^2 (\delta \mathbf{k}) \exp(i\delta \mathbf{k} \cdot \mathbf{r})[|\Psi(\delta \mathbf{k})\rangle_c + |\Psi(\delta \mathbf{k})\rangle_v]$. Thus, in order to localize graphene electrons on a single sub-lattice (A or B), one needs to superpose states at different energy, that is states from the valence band and from the conduction band, which arise from different valleys. Therefore each electronic wave function at fixed energy has components on both sub-lattices, possibly with equal weight. The only exception are states exactly at zero energy in the presence of a perpendicular magnetic field - their number increases to macroscopic size when a magnetic field is applied. In this case (the zero-energy level), states in the **K** valley reside on one sub-lattice (say B) and those in **K'** on the other one. The problem of graphene electrons localization mechanism on a single sub-lattice, other than that corresponding to the zero-energy level in the presence of a magnetic field, needs a serious investigation which we wish to take up in a future communication.

**REFERENCES**
[1]M. Sato, and S. Fujimoto, Phys. Rev. B 79 094504(2009).
[2] M. Sato, Y. Takahashi, and S. Fujimoto, Phys. Rev. Lett. 103 020401(2009); M. Sato, Y. Takahashi, and S. Fujimoto, Phys. Rev. B 82 134521 (2010).
[3]J.D. Sau, R.M. Lutchyn, S. Tewari, and S. Das Sarma, Phys. Rev. Lett. 104 040502(2010); R.M. Lutchyn, J.D. Sau, and S. Das Sarma, Phys. Rev. Lett. 105 077001(2010).
[4] E. Majorana, Nuovo Cimento 5 171(1937).
[5] D. Huertas-Hernando, F. Guinea, and A. Brataas, Phys. Rev. B **74**, 155426 (2006);Eur. Phys. J. Special Topics **148**, 177 (2007).
[6] H. Min, J. E. Hill, N. A. Sinitsyn, B. R. Sahu, L. Kleinman, and A. H. MacDonald, Phys. Rev. B 74, 165310 (2006).
[7] H. Zhang, C. Lazo, S. Blügel, S. Heinze,and Y. Mokrousov,Phys. Rev. Lett. 108, 056802 (2012).
[8] A. Varykhalov, J. Sanchez-Barriga, A.M. Shikin, C.Biswas, E. Vescovo, A. Rybkin, D. Marchenko, and O. Rader, Phys. Rev. Lett. 101, 157601 (2008).
[9] K. M. McCreary, A. G. Swartz, W. Han, J. Fabian, and R. K. Kawakami, Phys. Rev. Lett.109, 186604 (2012).
[10] D. Ma, Z. Li, and Z. Yang, Carbon 50, 297 (2012).
[11] A.A.Abrikosov, Sovjet Physics – JETP **5**, 1174 (1957).
[12] When graphene is doped with some magnetic atoms on the top of carbon atoms, apart from the magnetic proximity-induced exchange field and the interaction-induced Rashba spin-orbit coupling, the imbalanced AB sub-lattice potentials is also introduced [J. Ding, Z. H. Qiao, W. X. Feng, Y. G. Yao, and Q. Niu, Phys. Rev. B 84, 195444 (2011)].
[13] H. Suzuura, T. Ando, Phys. Rev. Lett. 89, 266603 (2002).
[14] D.V. Khveshchenko, Phys. Rev. Lett. 97, 036802 (2006).
[15] E. McCann, K. Kechedzhi, Vladimir I. Fal'ko, H. Suzuura, T. Ando, and B.L. Altshuler, Phys. Rev. Lett. 97, 146805 (2006); K. Kechedzhi, E. McCann, V.I. Fal'ko, H. Suzuura, T. Ando, and B.L. Altshuler, Eur. Phys. J. Special Topics 148, 39–54 (2007).
[16] B. Dora, M. Gulacsi, and P. Sodano, Phys. Status Solidi RRL 3, 169 (2009).
[17] Atiyah and Singer, Ann. of Math. 87, 485 (1968).
[18] F. Wilczek, "Majorana returns",Nature Physics 5, 614 (2009).
[19] P. Ghaemi, and F. Wilczek, Phys. Scr. T 146, 014019 (2012).
[20] C. L. Kane, and E. J. Mele, Phys. Rev. Lett. 95, 226801 (2005).
[21] D. Huertas-Hernando, F. Guinea and A. Brataas, Phys. Rev. B 74 , 155426 (2006).
[22] D. Huertas-Hernando, F. Guinea, and A. Brataas, Phys. Rev. Lett. 103 , 146801 (2009).
[23] J. Zhang, B. Zhao, and Z. Yang, Phys. Rev. B 88, 165422 (2013)).
[24] T. Yokoyama, Phys. Rev. B 87, 241409(R) (2013).
[25] L.Fu. and C. Kane, *Phys. Rev. Lett.* **100,** 096407 (2008). The experimental signatures of Majorana zero modes have been reported by these authors as zero bias peaks in tunneling spectroscopy of a single quantum wire with strong SO coupling which was coupled with an s -wave superconductor through proximity effect.

[26] J. R. Williams, A. J. Bestwick, P. Gallagher, S. S. Hong, Y. Cui, A. S. Bleich, J. G. Analytis, I. R. Fisher, and D. Goldhaber-Gordon, arXiv:1202.2323 (2012).

[27] M. T. Deng, C. L. Yu, G. Y. Huang, M. Larsson, P. Caroff, and H. Q. Xu, arXiv:1204.4130 (2012).

[28] }.L. P. Rokhinson, X. Liu, and J. K. Furdyna, arXiv:1204.4212(2012).

[29] M. Ezawa, New J. Phys. 14, 033003 (2012).

[30] M. Ezawa, Phys. Rev. Lett. 109, 055502 (2012).

[31] M. Ezawa, Phys. Rev. Lett. 110, 026603 (2013).

[32] M. Ezawa, cond-mat/arXiv:1205.6541.

[33] M. Ezawa, cond-mat/arXiv: 1207.6694.

[34] We have found, through the derivation of the total transmission probability in a normal-magnetic-normal silicene junction, that for the moderate 'M' the semiconductor-VSPM-band insulator transition is not destroyed, it only gets smeared out.

[35] P. Lounesto: *Clifford algebras and spinors*, London Mathematical Society Lecture Notes Series 286, Cambridge University Press, Second Edition (2001).

[36] Y.-J. Lin, R. L. Compton, K. Jimenez-Garcia, J. V. Porto, and I. B. Spielman, Nature 462, 628 (2009)..

[37] Y.-J. Lin, K. Jimenez-Garcia, and I. B. Spielman, Nature 471, 83 (2011).

[38] P. Wang, Z.-Q. Yu, Z. Fu, J. Miao, L. Huang, S. Chai, H. Zhai, and J. Zhang, Phys. Rev. Lett. 109, 095301 (2012).

[39] L. W. Cheuk, A. T. Sommer, Z. Hadzibabic, T. Yefsah, W. S. Bakr, and M. W. Zwierlein, Phys. Rev. Lett. 109, 095302 (2012).

[40] N. Goldman, G. Juzeliunas, P. Ohberg, and I. B. Spielman, ArXiv e-prints (2013), arXiv:1308.6533 [cond-mat.quant-gas].